\documentclass[sigconf, screen, nonacm]{acmart}

\AtBeginDocument{%
  }

\usepackage{wrapfig}
\usepackage{multirow}
\usepackage{xcolor}
\usepackage{colortbl} 
\newcommand{\method}{\texttt{STEALTHsense}}
\usepackage{amsmath}
\usepackage{mathtools}
\usepackage{amsthm}
\usepackage{mathrsfs}
\usepackage{subcaption}

\begin{document}

\title{Non-verbal Hands-free Control for Smart Glasses using Teeth Clicks}

\author{Payal Mohapatra*}
\affiliation{%
  \institution{Northwestern University}
  \city{Evanston}
  \country{USA}}
\email{payal.mohapatra@northwestern.edu}

\author{Ali Aroudi}
\affiliation{%
  \institution{Meta Reality Labs}
  \city{Redmond}
  \country{USA}
 }

\author{Anurag Kumar}
\affiliation{%
  \institution{Meta Reality Labs}
  \city{Redmond}
  \country{USA}
}

\author{Morteza Khaleghimeybodi}
\affiliation{%
 \institution{Meta Reality Labs}
  \city{Redmond}
  \country{USA}
  }

\newcommand\blfootnote[1]{%
  \begingroup
  \renewcommand\thefootnote{}\footnote{#1}%
  \addtocounter{footnote}{-1}%
  \endgroup
}

\begin{abstract}
Smart glasses are emerging as a popular wearable computing platform potentially revolutionizing the next generation of human-computer interaction. The widespread adoption of smart glasses has created a pressing need for discreet and hands-free control methods. Traditional input techniques, such as voice commands or tactile gestures, can be intrusive and non-discreet. Additionally, voice-based control may not function well in noisy acoustic conditions. We propose a novel, discreet, non-verbal, and non-tactile approach to controlling smart glasses through subtle vibrations on the skin induced by teeth clicking. We demonstrate that these vibrations can be sensed by accelerometers embedded in the glasses with a low-footprint predictive model. Our proposed method, called \method{}, utilizes a temporal broadcasting-based neural network architecture with just 88K trainable parameters and 7.14M Multiply and Accumulate (MMAC) per inference unit. We benchmark our proposed \method{} against state-of-the-art deep learning approaches and traditional low-footprint machine learning approaches. We conducted a study across 21 participants to collect representative samples for two distinct teeth-clicking patterns and many non-patterns for robust training of \method{}, achieving an average cross-person accuracy of 0.93. Field testing confirmed its effectiveness, even in noisy conditions, underscoring \method{}'s potential for real-world applications, offering a promising solution for smart glasses interaction.

\end{abstract}





\maketitle

\section{Introduction}
\blfootnote{*Work done during internship at the Meta Reality Labs, Redmond, USA.}

  There has been a meteoric rise in the popularity of smart glasses wearable technologies~\cite{waisberg2023meta} in recent years with wide adoption across industrial~\cite{aksut2024using}, medical~\cite{behmann2024visibility}, and daily living scenarios~\cite{mahato2024wearable} for a myriad of applications. Generally, smart glasses provide the ability to interact hands-free with various applications of the device and also enable virtual or augmented reality (VR, AR) experiences~\cite{survey2023}. In recent years, this immersive technology has become more influential with expanded capabilities like spatial audio, Artificial Intelligence (AI) assistant, seamless command over applications like music playback, receiving or declining calls  etc.~\cite{spatial-audio, mehra2021audio}. The verbal or limb-based conventional modes of the human-smart-glass interfaces are not discreet and prove intrusive in social settings. As a motivating example, consider a smart-glasses user engaged in an immersive exercise like rowing~\cite{castillo2024design} in a shared space which engages both their hands and they need to control the on-device music playback. This simple yet compelling example highlights the practical demand of exploring hands-free non-verbal communication to expand the usability of smart glasses. Furthermore, employing voice-based commands in a noisy or public environment not only lacks discretion and can be intrusive but also proves to be ineffective due to acoustic corruption. Voice-based commands are unsuitable for participants with speech disfluency~\cite{mohapatra2022speech, mohapatra2023efficient, mohapatra2024missingness}. 

\begin{figure}
    \begin{center}
    \includegraphics[width=0.9\linewidth]{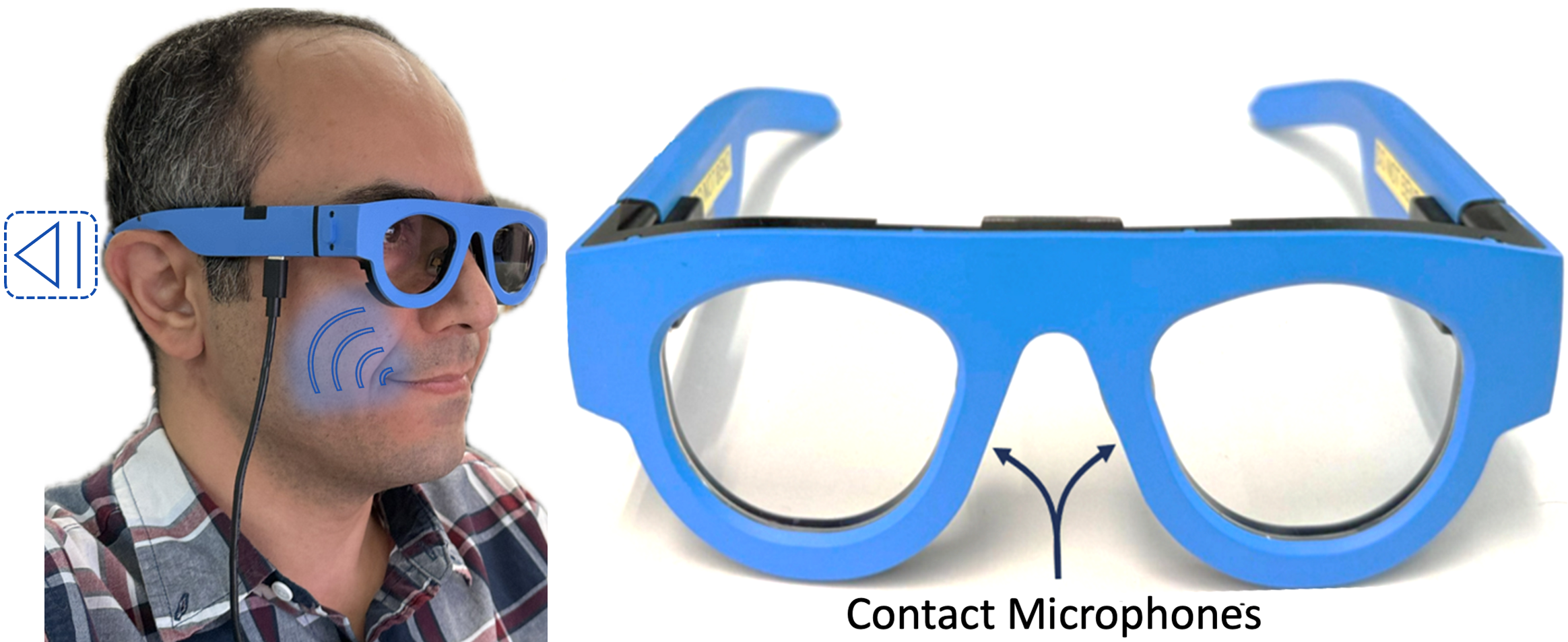}
    \caption{\method{} leverages the accelerometers embedded on the nose pads of the smart glasses to pick up non-vocal discreet teeth-clicking gestures for a seamless control interface using a lightweight real-time pattern recognition pipeline.}
    \label{fig:application_focused_overview}      
    \end{center}
\end{figure}

  Similarly, numerous AR/VR applications using smart glasses assert their utility in physiotherapy and rehabilitation programs~\cite{carlson2023virtual, canady2023virtual}, demanding adaptation to minimal limb involvement. Users with upper body or speech disability~\cite{simpson2008tooth} will be more inclusive and empowered with a technology option that does not require the use of voice or limbs.  Incorporating measures to ensure the inclusivity of minority users is a key pillar of accessible wearable technologies. Additionally, many health-focused applications like nursing education~\cite{nursing} and ergonomic correction~\cite{posture} can benefit from the unobtrusiveness of such a non-verbal hands-free control. Motivated by these factors, we explore a novel modality of user control for smart glasses using teeth-clicking gestures picked up by accelerometers on the nosepad of the smart glasses as shown in Figure~\ref{fig:application_focused_overview}. Such technology can also pave the way for non-biometric user authentication~\cite{xie2022teethpass, wang2022toothsonic} using discreet oral gestures. With a projection of 3.9 million unit sales of smart glasses by 2024 and an approximate anticipated revenue of 35 billion U.S. dollars by 2026~\cite{survey2023}, the exploration of this discreet real-time communication technology for AR glasses is exceptionally timely.
  
\begin{figure}[b]
\begin{center}
    \includegraphics[width=\linewidth]{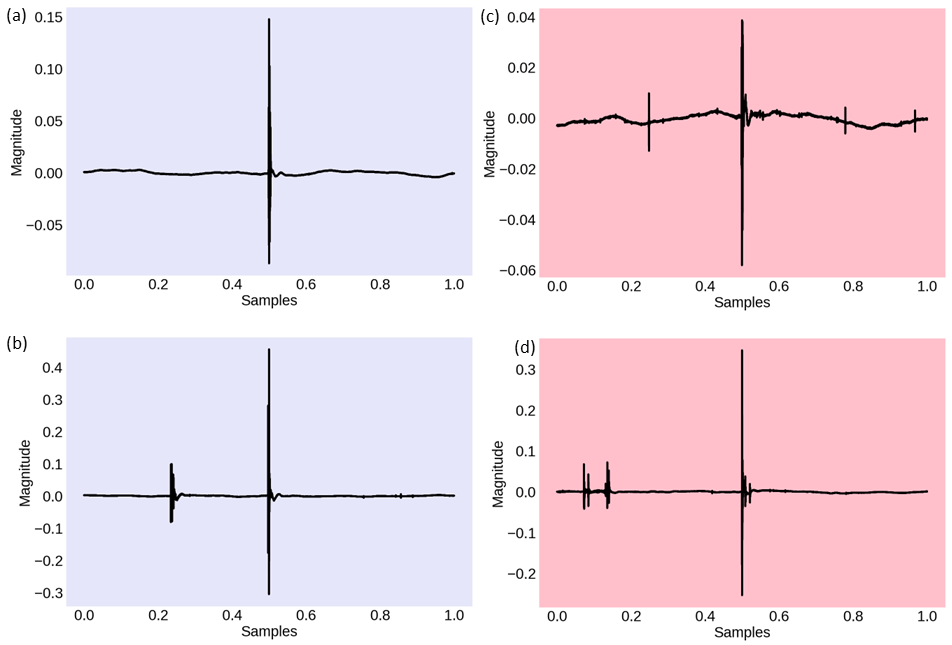}
\end{center}
\caption{Illustration of (a) an ideal template for single teeth click (pattern 1), (b) an ideal template for double clicks (pattern 2) and the corresponding (c) non-ideal pattern instance for single teeth click and (d) a non-ideal patterns instance for double teeth click due to variation in dental anatomy.}   
\label{fig:dental_anatomy}
\end{figure}

Although past works have delved into tooth-click as an interface in various designs ranging from behind-the-ear augmentation~\cite{simpson2008tooth}, earbuds~\cite{xie2022teethpass} to headbands~\cite{ashbrook2016bitey}; picking teeth-clicking signals using accelerometers placed on the nosepad of a pair of smart glasses has not been explored in the past. In this paper, we introduce \method{}, a highly performant novel discreet communication technique for smart glasses with thorough characterization. Additionally, most of the previous works detect teeth-clicking events in isolation (when they are not corrupted by motion, or background noise) against controlled settings; in contrast \method{} framework is designed to learn robust representations agnostic to artifacts, inspired by acoustic event detection techniques for generalization (see Section~\ref{sec:related_works}). Furthermore, comparing Figure~\ref{fig:dental_anatomy}. (a, b) and Figure~\ref{fig:dental_anatomy}. (c, d) respectively, undeniably highlights variations in the same teeth-clicking pattern across users due to diverse dental anatomies. Hence, there is no straightforward template that is universally applicable across participants and a more generalizable analytics framework is required.

There are several challenges in realizing our vision of seamless discreet interaction with smart glasses through teeth-clicks. Below we outline these challenges and our solutions to these problems, which also highlights some of the crucial contributions of this paper. 

\begin{figure}
    \begin{center}
    \includegraphics[width=0.8\linewidth]{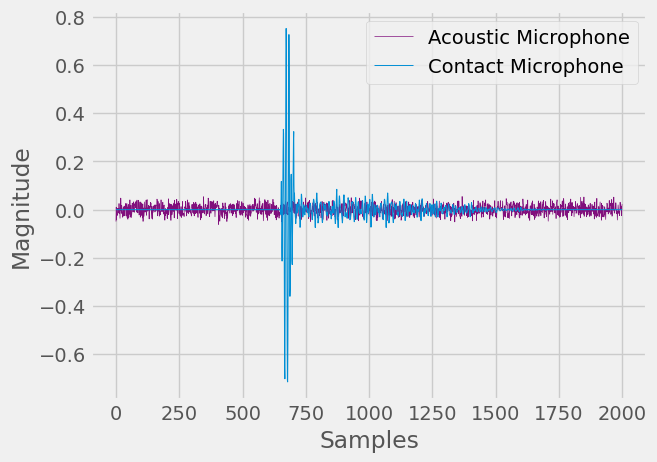}
    \caption{Motivating example to illustrate the simultaneous response for a single teeth click captured by an on-device nose-pad accelerometer and an acoustic microphone. Such nuanced discreet dental gestures can be picked up only through a nose-pad accelerometer modality. }
    \label{fig:acc_cm_motivation}      
    \end{center}
\end{figure}

\begin{itemize}
    \item \textbf{Signal capture and data}: The first major challenge is capturing the signal of interest that is teeth-clicks from the user or smart-glass wearer. Teeth-click signals are expected to be subtle (short-duration and of very low intensity) making sensors such as acoustic microphones mounted on smart glasses unsuitable for capturing teeth clicks. Moreover, the presence of noise and other signals in the environment may further prohibit uses of acoustic microphones for capturing high-quality teeth-clicks. To address this, we propose to use \textbf{nosepad-based accelerometers} to capture our signal of interest. Figure~\ref{fig:acc_cm_motivation} illustrates the difference in a teeth-click signal captured by an acoustic microphone and through our proposed system which uses nose-pad accelerometers as a contact microphone. This manuscript uses the terms nose-pad accelerometers and contact microphones interchangeably. We design a system, data collection protocol and annotation scheme to develop a dataset from 21 participants for two different types of teeth-clicking (single-click and double-click) patterns.

    \item \textbf{Detection algorithm}: Detecting teeth-clicks is also very challenging. Teeth-click signatures are person-specific depending on their dental patterns. Certain forms of oral conditions/disabilities can also impact the characteristics of the teeth-clicks. Moreover, we want our system to be able to detect different patterns of teeth-clicks which further introduces inter and intra-class variations. To add-on, certain activities such as chewing and speaking can also generate teeth-click-like signatures. We expect a usable system to be robust to such false patterns. We design \textbf{a novel neural network-based detection} approach and achieve high accuracy of \textbf{0.93} even on unseen users. 

    \item \textbf{Field testing}: A crucial consideration in the real-world system integration of such a system is its lightweight nature, ensuring a small compute footprint and low latency across diverse applications. We meticulously customized both the acoustic input features and the neural network architecture to facilitate deployment on smart glasses and real-time usage. Notably, the neural network comprises only \textbf{88K trainable parameters and approximately 7.1M multiply-and-accumulate units per second of inference}, ensuring feasibility in this regard.

\end{itemize}

We test the system with seen as well as unseen participants. Notably, we achieve \textbf{93\%} balanced accuracy when tested on data from unseen participants. Our exhaustive experiments provide several insights into the design of such a system, e.g. the choice of input features (Figure~\ref{fig:feat_imp}), task-specific augmentations for robust representation (Figure~\ref{fig:robustness}), etc. Moreover, our results are supported by field testing performance with a score of 3.72 out of 5 for adoption and 88\% users found the current prototype very accurate (score of 3 or more out of 5).

\noindent\textbf{Terminology.}
Throughout this manuscript, we refer to teeth-click as the act of clicking the upper and lower jaws rapidly with a closed mouth in an arbitrary way as a user prefers. It is more impulsive than teeth-grinding (commonly encountered during chewing). Doing a click once is referred to as \textit{single click}, doing so twice is \textit{double click}, and so on. The duration between clicks in patterns with more than one teeth-click event is non-prescriptive. A user may use any combination of teeth pairs from the upper and lower jaws to issue a teeth-click as convenient. 

\section{Related Works}\label{sec:related_works}

\noindent \textbf{Teeth-clicking Interfaces.} One of the earliest works to use tooth-click for human-computer interaction by Simpson et al.~\cite{simpson2008tooth} controls mouse button clicks for disabled participants using a 3-axis accelerometer behind the ears. Another work~\cite{ashbrook2016bitey} explored the glasses form factor with an additional head-band unit tying the legs of the glasses and containing a bone-condition microphone that perches just above the ears. Their goal is to identify different types of teeth-clicking based on the choice of the pair of teeth. However most if not all of the past works ~\cite{nguyen2018tyth, simpson2008tooth, ashbrook2016bitey} isolate the teeth-clicking event and conduct experiments in controlled settings to demonstrate the efficacy of their method, which will not translate well in the world where robustness against speaking, chewing, motion artifacts, etc. is imperative. One of the recent works by Xie et al.~\cite{xie2022teethpass} proposes in-ear bone binaural earbuds to detect dental occlusions and have considered the impact of speech scenarios and motion artifacts in their design in the context of user authentication. Different from the past works we use a new instrumentation to pick up subtle vibrations due to teeth clicking through the nose and design a sophisticated low-overhead generalizable learning pipeline to perform consistently across users under varying artifact scenarios.

\noindent \textbf{Algorithmic Considerations.} A well-known class of algorithms for classifying acoustic events is Audio event detection~\cite{kumar2016audio, mesaros2017detection}. A handful of works~\cite{lea2022nonverbal, chabot2021detection} have considered the scenario of detecting or classifying audio events based on non-verbal sounds. For example, Lea et al.~\cite{lea2022nonverbal} aims to improve the voice assistance experience for users with speech disfluency by recognizing non-verbal sounds. One common denominator in most of the non-verbal sound detection works~\cite{igarashi2001voice, funk2020non, lea2022nonverbal} is the primary use of an acoustic microphone for data acquisition. Our design proposes the use of a nose-pad accelerometer which is relatively immune to acoustic background noise but comes with its challenges of susceptibility to motion artifacts, skin contact, etc. Some other works~\cite{chabot2021detection, xie2022teethpass} explore non-verbal body sounds in the context of in-ear hearable devices which may isolate the user from the ongoing acoustic scene. However, our design features an accelerometer on the nose pad of the smart glasses which is a more complacent placement for regular continuous use. Moreover, the past works feature a closed set classification case which may not be robust to variations of target and non-target classes. However, we consider event patterns like teeth click pattern one vs. teeth click pattern two in addition to no event case [or to non-target events]. The purview of the no-event class is non-exhaustive, so the key focus during training our model is to learn robust target pattern representation. 

\noindent \textbf{Discreet Gesture Interfaces in Wearables.}
There is a longstanding inclination to command smart devices discreetly and unobtrusively. Several researchers investigate unobtrusive techniques with subtle gestures like wearable hand gestures~\cite{ashbrook2011nenya, xu2022enabling, pasquero2011haptic}, haptic feedback through earbuds~\cite{xu2020earbuddy,zhao2007earpod}, eye gaze based control~\cite{tuisku2012wireless, chin2008integrated, zhao2023robotic}, silent speech using orally embedded sensors~\cite{sahni2014tongue} or using necklace~\cite{zhang2021speechin}, etc. As previously mentioned, while tooth-clicking is a preferred method for issuing discreet commands, the feasibility of detecting such signals from accelerometers embedded in the nosepad of glasses has not been explored before. In contrast to our prior work~\cite{khaleghimeybodi2022system}, which focused on capturing non-verbal cues from the back of the ear, in this paper we highlight the effectiveness of utilizing nose-pad accelerometers for recognizing teeth-clicking cues.



\section{System Design} \label{sec:sys_design}

In this section, we offer a comprehensive overview of \method{} and a formal problem statement. Initially, we outline our custom dataset curation procedure, delineating the specifics of our data collection protocol, comprehensive data analysis, characterization methods, and presenting relevant data statistics. Next, the underlying neural network is described with a thorough examination of our design choices, encompassing feature selection, data augmentation strategies, and architectural innovations. These choices collectively contribute to our notable achievement of an overall balanced accuracy of 0.93 when applied to previously unseen participant data.

\begin{figure*}[t]
  \centering
  \includegraphics[width=0.9\linewidth]{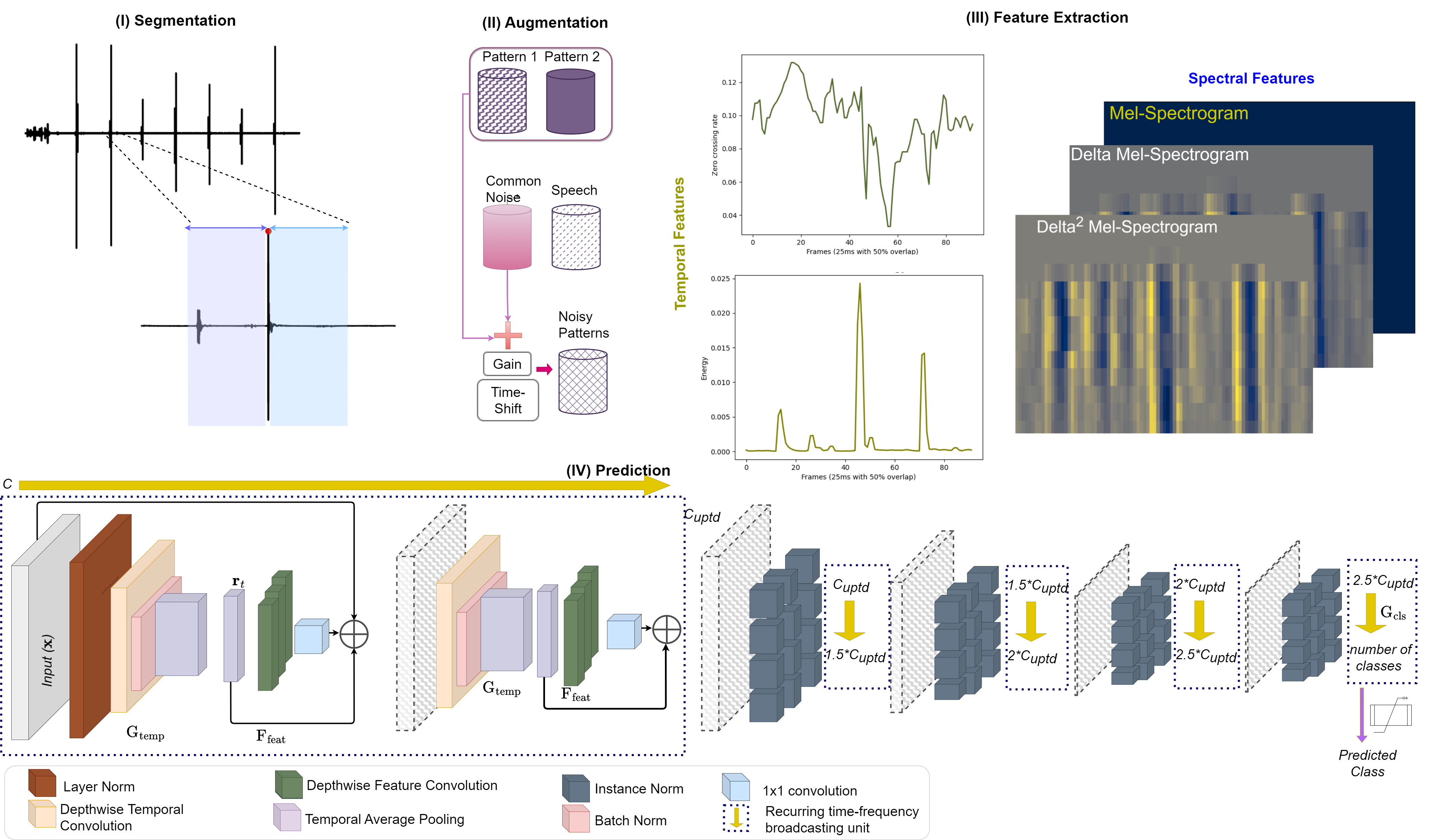}
  \caption{Overall System Architecture illustrating I) segmentation of patterns using an annotator model, II) data augmentation using gain, time shifting, and additive noise from the pool of common noise using properties from the SNR study, III) feature engineering to combine spectral and temporal properties of the signal and IV) predictor network architecture for detection the event. }


 \label{fig:algoDesign}
\end{figure*}

\noindent\textbf{Overview.} \method{} is trained using a custom dataset corpus from 21 participants and annotated using a temporal-thresholding and rule-based algorithm to serve as ground truth. The dataset consists of two distinct tooth-click patterns and several negative instances including self-speech, mastication events, motion, etc. Next, we characterize our signal of interest and design a tailored augmentation pipeline for nose-pad accelerometer data. Finally, we develop a low-footprint neural network model robust to inter-person variability and noise corruption and capable of real-time inference. The overall \method{} system is illustrated in Figure~\ref{fig:algoDesign}.

\noindent\textbf{\method{} Hardware.} Our hardware consists of a custom-built smart glasses developer platform equipped with a Khadas VIM3 Amlogic A311D compute unit. The accelerometer sensors (VPU 14DB01A) are located within the nose pad of the glasses prototype and are used to collect 2-channel vibration data driven by the teeth clicks as highlighted in Figure~\ref{fig:application_focused_overview}. In all our analyses, data from these dual-channel accelerometer streams are averaged and converted into a single-channel and then, this 1D information is used for post-processing. The form factor of the system allows full mobility for the user. The overall build of the glasses prototype is similar to previous designs in~\citet{mehra2020hybrid, anderson2023multimodal}, etc. The user wears the glasses prototype and issues teeth-clicks, which are picked up by the nose-pad accelerometers. The information captured by these nose-pad accelerometers then goes through the inference engine and based on the predictor output, it carries out an application control. As an illustrative use case, we have implemented a music playback control through this interface for real-time demonstration. Figure~\ref{fig:application_focused_overview} illustrates the overview of the system and Figure~\ref{fig:algoDesign} dives into the inference framework. 

\noindent\textit{Problem Definition.} Our objective is to design a lightweight classifier capable of real-time inference on the device to detect two distinct teeth-clicking patterns from the nose-pad accelerometers in the smart glasses robustly. Our lens of robustness is twofold; 1) robust to inter-person variation and provide high performance to unseen users, and 2) robust to non-teeth click patterns (acoustic noise, user movements, self-speech, and other mastication actions).

\subsection{Dataset Collection}\label{sec:data}

\subsubsection{Study Protocol} 
We collected data from a cohort of 21 multilingual participants, 7 female and 14 male individuals, aged from 23 to 59 years old. Some of our participants had tooth fillings (8 participants) and some (5 participants) had their wisdom teeth removed, providing a diverse teeth-anatomy in our data pool. The study protocol involves collecting about 10 examples for each type of target pattern from every participant. We collect data for two different types of teeth-clicking patterns (pattern 1 or single-clicks and pattern 2 or double-clicks). Our study design consists of a visual aid that prompts the participants to issue different patterns of teeth-clicking gestures. Such a predefined guide allows us to enable time-based thresholding (refer to Section~\ref{subsubsec:annotate}) in the data-annotation pipeline. Our approach intentionally avoids highly prescriptive guidelines regarding gesture execution, refraining from specifying detailed parameters such as issuing a single click with molars or enforcing restrictions on unilateral tooth engagement. This deliberate choice aims to emulate real-world user scenarios characterized by general instructions, fostering a more ecologically valid experimental environment.

Additionally, to make the model resilient to self-speech, subject-specific speech samples for the no-pattern class are also collected for each individual. The participants are instructed to read out a set of Harvard sentences~\cite{licgee1969w} for this scenario. Our experimental protocol is approved by the Institutional Review Board (IRB).

In addition, we also collected a diverse set of common non-teeth-click-pattern examples, encompassing elements such as background music, babble noise, motion artifacts manifested through activities such as walking and nodding head, as well as instances of chewing, drinking, and periods of silence. We further characterize the impact of background acoustics on the nose-pad accelerometers in Section~\ref{subsubsec:aug} and identify their scope of impact on our device and incorporate the results suitably into our design pipeline. This comprehensive collection serves to broaden the scope of our dataset, capturing a spectrum of non-teeth-click patterns encountered in real-world scenarios. Note that we use the notation single click and pattern 1 interchangeably. Similarly, we use double click and pattern 2 interchangeably.

\subsubsection{Annotation Framework}\label{subsubsec:annotate}
Before we look at the annotation schema for segmenting the continuously recorded event patterns, we explain the choice of design specifics in terms of spectral analysis and preprocessing, and target event length.

\noindent\textbf{Empirical Target Event Length Estimation.} The frame length is determined based on a heuristic estimation of the longest period observed to capture the target patterns (complete single click or a double click). 
The empirical duration for a single and double teeth click is approximately 15ms and 500ms respectively to ensure complete occurrence of an event within every segmented audio.

\noindent\textbf{Spectral Analysis and Preprocessing.} Oral occlusal movements picked up by a nose-pad accelerometer are not well studied, so we conduct a spectral analysis to arrive at the cut-off frequencies for filtering the signal of interest. We leverage pilot analysis data to extract pattern 1 (single click events) from 3 participants with a fixed frame length of 25ms. We utilize the non-event segment of the file to determine a noise floor. The only preprocessing done here is notch filtering (60Hz and harmonics) to get rid of electrical noise. For Participant 1 the resonant frequency range is around 500 Hz - 800 Hz and for Participants 2 and 3 it's around 5kHz as shown in Figure~\ref{fig:interperson}. (a-c). From this study, we arrive at the lower and upper cut-off for the bandpass filter (BPF) to be between 300 Hz to 5kHz.  Also, the frequency characteristics indicate a strong variability between subjects for the same pattern as shown in Section~\ref{subsub:feasibility}.

\noindent\textbf{Annotator Model.} The continuous data streams for each participant need to be annotated and segmented. We use notch filtering with a cut-off frequency of 60 Hz and three subsequent harmonics followed by band-pass filtering with a cut-off frequency of 300 Hz and 5kHz. Based on a fixed time-based thresholding (events occur at least 5 seconds apart based on the visual cues in the experimental protocol) and a peak-prominence threshold~\cite{fejes2008findpeaks} we determine the local maximum peak. This is sufficient if we were to detect only a single-click pattern. However, since we need to extract fixed segments containing complete events of patterns consisting of more than one-teeth click, we adopt a strategy to extract a fixed number of samples before and after the detected peak as shown in Figure~\ref{fig:algoDesign}. (I). This results in 1s length for the segmented audio since for accommodating a double click event we need 500 ms. The non-pattern data streams are simply segmented to 1 s length for uniformity. Overall, we obtain three classes (pattern 1, pattern 2, and no pattern) for each participant. The data statistics are shown in Table~\ref{tab:data_stat}.

\begin{table}
\centering
\caption{Summary of the data statistics for \textit{target patterns} and representative \textit{no pattern} samples collected from 21 participants. Note that, throughout this manuscript, we use the notation single click and pattern 1 interchangeably. Similarly, we use the notation double click and pattern 2 interchangeably.} 
\begin{tabular}{ccc}
\hline
\multicolumn{1}{c}{\textbf{Class}} & \multicolumn{1}{c}{\textbf{Data Details}}                                   & \multicolumn{1}{c}{\textbf{Samples}} \\ \hline
\multirow{6}{*}{No Pattern}            & \begin{tabular}[c]{@{}c@{}}Speech \\ (Subject Specific)\end{tabular}   &  840                                   \\ 
                                     & Chewing         & 60                                      \\  
                                     & Motion Artifact & 45                                       \\ 
                                     & \begin{tabular}[c]{@{}c@{}}Background Accoustic\\  Babble Noise\end{tabular} & 30          \\  
                                     & Background Music & 20                                       \\  
                                     & Silence  & 180                                       \\ \hline
Pattern 1                            & Single Teeth Click  &    343                                   \\ \hline
Pattern 2                            & Double Teeth Click  &    381                                   \\ \hline
\end{tabular}\label{tab:data_stat}
\end{table}

\subsection{\method{} Predictor Framework} \label{subsec:Algo_Design}
A custom dataset is curated as per Section~\ref{sec:data}, to train a low-footprint deep learning framework to identify the positive teeth-clicking gestures. The following sections describe a tailored data augmentation pipeline unique to \method{}, the input features, and the overall network design and training details.

\subsubsection{Data Augmentation}\label{subsubsec:aug}
Nose-pad accelerometers are known to be resistant to background noise compared to acoustic microphones~\cite {drugman2020audio, hiraki2023external}. However, some empirical studies~\cite{tagliasacchi2020seanet} do indicate that the acoustic properties of the signal picked up by the nose-pad accelerometer aid in speech enhancement. To learn robust representations in real-world scenarios, we need to account for this kind of corruption. We will present our characterization of the impact of background acoustic noise on the signal captured by the nose-pad accelerometer in our case.

Our data collection protocol described in the previous section is carried out under controlled settings (minimal movement and background noise) to facilitate a true ground truth annotation. However, to allow robust performance in the real world we need to expose \textit{noisy} samples to train the predictor network. We carry out three types of data augmentation: additive noise, signal gain (-6dB to +6dB), and temporal shift (circular - samples shifted to the right are appended from the left) for the target patterns. 
Given the uniqueness of our application, we cannot leverage additive noise from AED benchmarks~\cite{dubey2022icassp}. We first characterize the impact of the noise floor in the signal. 

\begin{figure}
\begin{center}
    \includegraphics[width=\linewidth]{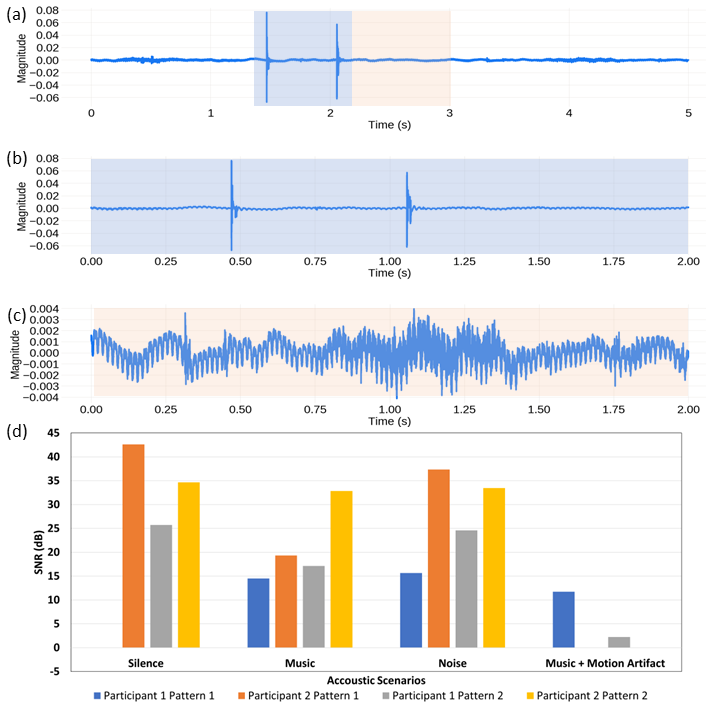}
    \caption{Pilot study on SNR characterization showing (a) original waveform captured in a sound room with a signal-plus-noise segment (blue highlight) and a noise segment (red highlight), (b) the trimmed segment corresponding to signal-plus-noise segment and (c) the noise segment. The duration for signal-plus-noise and noise segments are maintained to be equal. (d) SNR for two teeth-click pattern (single click and double click) for two participants.}
 \label{fig:snr_char}
\end{center}
\end{figure}

\smallskip \noindent \textbf{Characterizing the Impact of Noise floor.} To formally establish the most deterrent factors for a robust representation we need to train the model under real-world settings where there is acoustic background noise. Since collecting data in noisy settings hinders us from ground truth annotation, we augment noisy data by mixing acoustic background noise synthetically. Before doing this we need to characterize the impact of the background noise signal collected by the nose-pad accelerometer. This allows us to empirically come up with a SNR range for noise mixing. We collected event data from two participants in a moderately reverberant room with 1) surround music, and 2) babble noise. Figure~\ref{fig:snr_char}. (a) shows an example of such collected data consisting of teeth clicks and background noise. The segments of the signals where teeth click occurs are denoted as the 'signal plus noise' segments, while the intervals consisting of only background noise are denoted as the noise segments as shown in Figure~\ref{fig:snr_char}. (a). The only preprocessing we carry out is notch filtering to remove electrical noise. We compute the signal-to-noise ratio of a given audio segment as 
\[SNR_{db} = 10 \log 10((P_y - P_n)/P_n\]
where P$_y$ is the power of the signal-plus-noise segment and P$_n$  is the power of the noise segment. Note that the duration for signal-plus-noise and noise segments are maintained to be equal. We can draw few insights from this study. First, the overall SNR is positive indicating that background noise does not severely corrupt the target signals. Second, the maximum difference in SNR of background noise vs. silence is (for Pattern 1 Participant 2), ~23dB (refer Figure~\ref{fig:snr_char}.(b)). We use this as the specification of -23dB to +23dB from the noise pool of acoustic and motion artifact samples for mixing with clean pattern data offline to generate realistic \textit{tough} samples for training our model. More evaluations under various stages of noise mixing are discussed in Figure~\ref{fig:robustness}. 

\subsubsection{Feature Extraction}\label{subsubsec:feat_extraction}
Following up from the Section~\ref{subsubsec:annotate}, all data are segmented to 1s length and notch-filtered with the harmonics of 60 Hz using a second-order Butterworth~\cite{butterworth1930theory} band-stop filter infinite impulse response filter and a sampling rate of 48kHz, followed by bandpass filtering with cut-off frequencies at 300 Hz and 5kHz. 

We extract 13 log-Mel spectrogram features per input segment with 25ms frame and 50\% overlap between consecutive frames. Given the impulsive signature of our event of interest, intuitively features that capture the rate of change are well-suited for this application. Additionally, we extract the first and second derivatives of the log-mel features~\cite{wrede2003dynamic}. Apart from spectral features, we also compute two temporal features - zero-crossing rate (ZCR) and short-term energy (STE) as $ \sum_{t_0}^{t_{25}} |s(t)|^2 $ where $ s(t) $ is the streaming input and $ {t_0} $ to $ {t_{25}} $ denotes the frame duration of 25ms (translates to 1200 samples in this case). This results in a 41-dimensional feature set. Figure\ref{fig:algoDesign}.(III) shows a visualization of the features. More analysis on the impact of various features on the model performance is given in Section~\ref{subsec:feat_imp}.

\subsubsection{Predictor Network Design}\label{subsubsec:predictornw}
We aim at designing a lightweight network that is robust to self-speech, motion artifacts, and background acoustic noise to detect the teeth-clicking cues accurately. We categorize all the non-teeth click cues as one class, one teeth click also referred as pattern 1 and two consecutive teeth click also referred as pattern 2 as the other two classes, as shown in Table~\ref{tab:data_stat}. 

Past literature has shown the efficacy of depth-wise convolutions~\cite{howard2017mobilenets} and a broadcasting residual~\cite{kim2021broadcasted} for data and resource-constrained applications. We leverage the design proposed by Kim et. al~\cite{kim2021broadcasted} for efficient keyword detection in our application. The central idea of this architecture is repeated pooling of feature set to 1-dimension(D) and then broadcasting back to 2-dimension by using residual identity connections. This architecture was originally designed for speech applications with a homogeneous spectral feature set like mel-frequency-cepstral-coefficients which means it is beneficial to pool features to translate the data into purely temporal dimension. However, our application has unique attributes that justify using a mix of various spectral and temporal dimensions into account as described in Section~\ref{subsubsec:feat_extraction} and illustrated in Figure~\ref{fig:algoDesign}.III. We observe that instead of conducting pooling along the feature axis and then broadcasting along the temporal axis; the vice-versa operation performs better. Juxtaposing the two broadcasting techniques, temporal to feature-wise vs. feature-wise to temporal (as proposed originally by Kim et. al~\cite{kim2021broadcasted}) we observe performance on unseen participants as test data as 0.93 and 0.90 respectively.  We reason that given the heterogeneous nature of our input features the uniform pooling along time axis corrupts the discriminative features. 

\noindent \textbf{Architecture Design.} Our architecture consists of a recurring core learning unit for time-frequency broadcasting as shown in Figure~\ref{fig:algoDesign}. IV. We apply layer norm~\cite{srivastava2022conformer} to the input, $\mathbf{x} \in \mathbb{R}^{1 \times T \times F} $, where $T$ is the temporal length after feature extraction and $F$ is the number of extracted features to account for real-time varying data statistics. This is followed by a depth-wise temporal encoder, $\mathrm{G}_\mathrm{temp}$ implemented using 1D convolutions along the temporal axis. Since we use a different formulation to broadcast between time-frequency dimensions, sub-spectral normalization~\cite{chang2021subspectral, kim2021broadcasted} is not beneficial to our design, and the use of batch-normalization works well. The temporal convolution is followed by average pooling along the temporal dimension to obtain a 1D feature map following, $\mathrm{G}_\mathrm{temp}:\mathbb{R}^{1 \times T \times F}\xrightarrow{} \mathbb{R}^{C \times 1 \times F}$. The output from $\mathrm{G}_\mathrm{temp}$, $\mathbf{r}_{t}$ is normalized per instance as $\widehat {\mathbf{r}}_{t} = \frac{\mathbf{r}_{t} - \mu(\mathbf{r}_{t})}{\sqrt{\sigma(\mathbf{r}_{t})^2 + \epsilon}} \cdot \gamma + \beta$ where $\mu$ and $\sigma$ are the mean and standard deviation across the batch of training examples,  $\gamma$ and $\beta$ are learning parameters for each time step and $\epsilon$ is a small constant added for numerical stability. The normalized 1D representation, $\widehat {\mathbf{r}}_{t}$, is given to a feature-wise encoder with a broadcasting unit, $\mathrm{G}_\mathrm{feat}: \mathbb{R}^{C \times 1 \times F}\xrightarrow{} \mathbb{R}^{C' \times T' \times F'}$. The key broadcasting operations to convert the 1D features to 2D are given in Equations~\ref{eqn} alternating based on the operative layer. 

\begin{equation} 
  \mathbf{r} = \begin{cases}
  \mathbf{x} + \mathrm{G}_\mathrm{temp}(\mathbf{x}) +  \mathrm{G}_\mathrm{feat}(\mathbf{r}_t) \\
  \mathrm{G}_\mathrm{temp}(\mathbf{x}) +  \mathrm{G}_\mathrm{feat}(\mathbf{r}_t)
  \end{cases}
\end{equation}\label{eqn}

Also, note that channels are updated after the temporal and feature-wise encoder steps but the 2D dimensions are consistent to allow seamless broadcasting. The Figure~\ref{fig:algoDesign}.IV illustrates the scaling along the channels at every recurring time-frequency broadcasting block. Finally, a classifier head, $\mathrm{G}_\mathrm{cls}$ applies softmax operation to its output, $\mathbf{r}_{cls} \in \mathbb{R}^{1 \times 1 \times C_{fin}}$, where $C_{fin}$ is the number of output classes to optimization of a cross-entropy objective given as,  
$$
\frac{1}{N_B}\sum_{i=1}^{N_B} \mathbf{r}_{fin} \log{\mathrm{G}_\mathrm{cls}(\mathbf{r})},
$$
where $N_B$ is the number of examples in a batch. This parameterization of channels within each time-frequency broadcasting unit can be configured to control the model size. Figure~\ref{fig:model_size} illustrates the impact of model size on performance. 


\subsubsection{Training}
Our current setup naturally suffers from a few data-related challenges: 1) imbalanced class distribution - pattern 1 and pattern 2 are at least 3 times smaller in size (refer Table~\ref{tab:data_stat}), 2) interperson variability and 3) modest dataset size.
We overcome the challenge of imbalanced classes using a sampler to re-weight the sampling weight per class in every minibatch. The augmentation as described in the previous section is applied with a probability of 0.7 to help with model generalization and increasing the dataset variety. We use a mix of spectral and temporal features which outperform the mere use of spectral features (even with a much higher resolution, refer Figure~\ref{fig:feat_imp}). We address the issue of inter-person variability by carefully crafting our representation learning pipeline with application-specific components, to overall increase the upper bound of generalization performance~\cite{ben2006analysis}. We use ADAM optimizer~\cite{kingma2014adam} to find minimize cross-entropy~\cite{good1952rational} loss function and early exit based on validation loss from in-domain samples(data held out from the training set participants) for model selection. Given our modest model and dataset size, hyperparameters of batch size and learning rate of 128 and 1e-3 worked best for this application. We use a batch size of 128. We used Pytorch~\cite{paszke2019pytorch} framework and trained on NVIDIA Tesla V100-SXM2 GPU. 

\section{Experimental Results}
We present the results of \method{}'s predictive abilities by first justifying our evaluation setup of non-overlapping users and its practical significance by answering the question - \textbf{Is there any inter-person variability?} 

\begin{figure}[h]
    \begin{center}
    \includegraphics[width=0.5\textwidth]{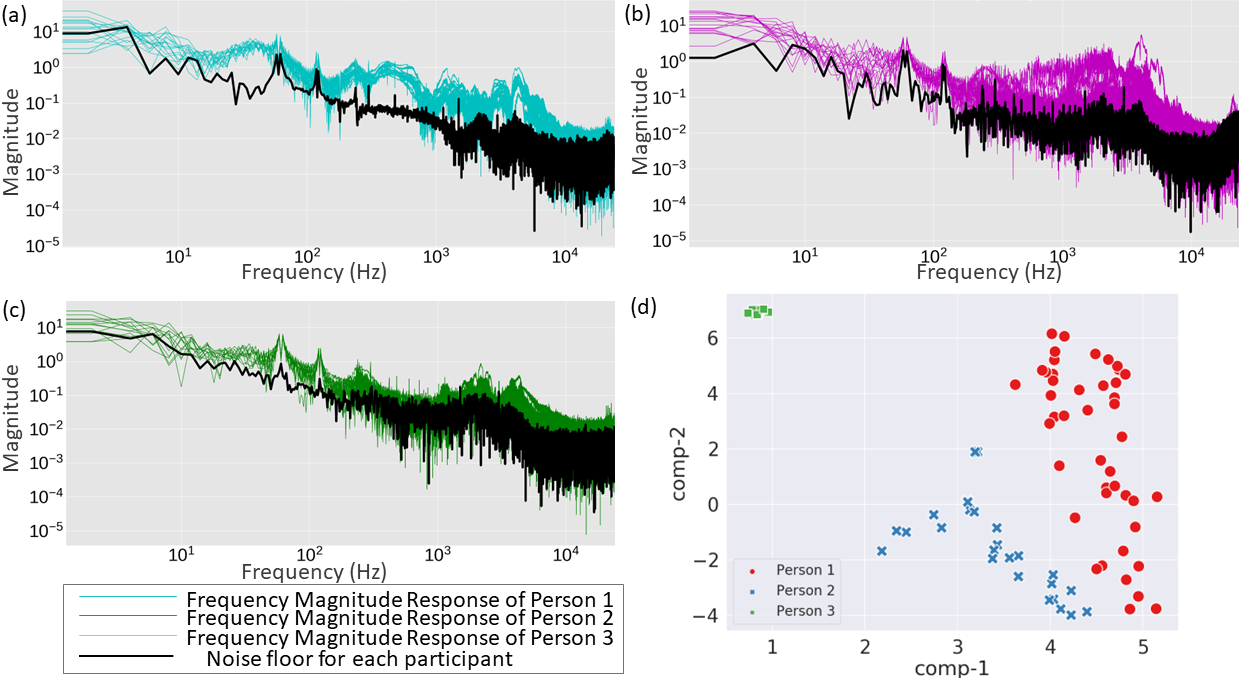}
    \caption{Illustration of spectral characteristics for single teeth click (pattern 1) with the corresponding noise floor for (a) participant 1 (resonant frequency range is around 500 Hz - 800 Hz), (b) participant 2 (resonant frequency range is around 5 kHz) and (c) participant 3  (resonant frequency range is around 5 kHz). (d) t-SNE projection for spectral and temporal features extracted from 3 participants for single teeth click pattern. These plots provide some evidence of inter-person variability.}
    \label{fig:interperson}   
    \end{center}
\end{figure}

We analyzed approximately sixty single teeth click patterns for 3 participants. These events are 1 second in duration. We do this analysis using two tools - 1) juxtaposing the individual magnitude responses of the Fourier transform of each sample, 2) viewing the clusterability of the samples as shown in Figure~\ref{fig:interperson}.  After conducting a spectral analysis for the single teeth-click (pattern 1) samples from each participant with the respective noise floor, we observe the signal of interest lies in different frequency bands for each participant as highlighted in Figure~\ref{fig:interperson}.(a-c) ranging from approximately 300 Hz to 5 kHz. Next, we plot the t-SNE projection in Figure~\ref{fig:interperson}.(d) for visualizing how the features for single teeth click (pattern 1) translate across individuals. We can observe that there is sufficient clustering to facilitate learning of a decision boundary for identifying the individual who is generating teeth click (also called person identification~\cite{mohapatra2023person}). To further validate this observation, we train a small classifier with data from three participants. The results show 0.94 accuracy in person identification merely from teeth-click signatures for data from 3 participants. We simply use the person identifier as the target label and train a Support Vector Classifier (SVC)~\cite{cortes1995support} with the linear kernel (for 3 classes in this case - person 1, person 2, and person 3) for this person-identification from teeth-click exploration. These results allude to various discriminative attributes for every person issuing the same teeth click and make our task more complex to generalize across unseen participants in the real world. This compels us to evaluate our design on participants without non-overlapping the data we use for training the predictor model.

\begin{table*}[]
\centering
\caption{Summary of various choices for predictor network performance. The \method{}'s framework is indicated with an asterisk and the best performance across all the settings is highlighted in \textbf{bold}.} 
\vspace{-10pt}
\label{tab:summar_res}

\resizebox{\textwidth}{!}{%
\begin{tabular}{l|cc|cc|cc|cc|c|c}
\hline
 &
  \multicolumn{2}{c|}{\textbf{Balanced   Accuracy}} &
  \multicolumn{2}{c|}{\textbf{\begin{tabular}[c]{@{}c@{}}No Pattern \\ F1 Score\end{tabular}}} &
  \multicolumn{2}{c|}{\textbf{\begin{tabular}[c]{@{}c@{}}Pattern 1\\ F1 Score\end{tabular}}} &
  \multicolumn{2}{c|}{\textbf{\begin{tabular}[c]{@{}c@{}}Pattern 2\\ F1 Score\end{tabular}}} &
   &
   \\ \cline{2-9}
\multirow{-2}{*}{\textbf{Model}} &
  \multicolumn{1}{c|}{Clean} &
  Noisy &
  \multicolumn{1}{c|}{Clean} &
  Noisy &
  \multicolumn{1}{c|}{Clean} &
  Noisy &
  \multicolumn{1}{c|}{Clean} &
  Noisy &
  \multirow{-2}{*}{\textbf{\begin{tabular}[c]{@{}c@{}}Multiply Accumulate(M) \\ (per second of input) \end{tabular}}} &
  \multirow{-2}{*}{\textbf{Model Size}} \\ \hline
SVC &
  \multicolumn{1}{c|}{0.76} &
  0.60 &
  \multicolumn{1}{c|}{0.94} &
  0.83 &
  \multicolumn{1}{c|}{0.65} &
  0.44 &
  \multicolumn{1}{c|}{0.70} &
   0.54 & -
   & -
   \\ 
XGBoost &
  \multicolumn{1}{c|}{0.74} &
  0.66 &
  \multicolumn{1}{c|}{0.97} &
  0.88 &
  \multicolumn{1}{c|}{0.58} &
  0.49 &
  \multicolumn{1}{c|}{0.66} &
   0.61 & -
   & -
   \\ 
ConvLSTM &
  \multicolumn{1}{c|}{0.89} &
  0.79 &
  \multicolumn{1}{c|}{0.59} &
  0.56 &
  \multicolumn{1}{c|}{0.32} &
  0.30 &
  \multicolumn{1}{c|}{0.67} &
  0.67 &
  0.91 &
  20155 \\ 
Broadcasting along Feature axis &
  \multicolumn{1}{c|}{0.90} &
  0.90 &
  \multicolumn{1}{c|}{0.96} &
  0.96 &
  \multicolumn{1}{c|}{0.89} &
  0.87 &
  \multicolumn{1}{c|}{0.86} &
  0.84 &
  7.58 &
  81143 \\ 
Broadcasting along Temporal axis* &
  \multicolumn{1}{c|}{\textbf{0.93}} &
  \textbf{0.91} &
  \multicolumn{1}{c|}{\textbf{0.98}} &
  \textbf{0.97} &
  \multicolumn{1}{c|}{\textbf{0.91}} &
  \textbf{0.89} &
  \multicolumn{1}{c|}{\textbf{0.90}} &
  \textbf{0.88} &
  7.14 &
  88687 \\ 
Broadcasting along Temporal axis + Attention &
  \multicolumn{1}{c|}{0.92} &
  0.85 &
  \multicolumn{1}{c|}{0.97} &
  0.96 &
  \multicolumn{1}{c|}{0.89} &
  0.76 &
  \multicolumn{1}{c|}{0.88} &
  0.80 &
  8.20 &
  248287 \\ \hline
\end{tabular}%
}
\end{table*}

\subsection{\method{}'s Prediction Performance}

We benchmark our proposed approach against other existing low-footprint methods from traditional computation to deep learning techniques. We use the standard composite classification metrics for evaluating our model's performance - balanced accuracy~\cite{kelleher2020fundamentals, brodersen2010balanced},  confusion matrices, and F1 score defined as, 
$$ F1_{score} = \frac{TP}{TP + \frac{1}{2}(FP + FN)},$$
where TP is True Positive, FP is False Positive and FN is False Negative. 

\noindent\smallskip\textbf{Machine Learning Classifiers} Support Vector Classifiers (SVC)~\cite{cortes1995support} are powerful engines that learn decision boundaries for high-dimensional data using only a few parameters. Previous works support the superiority of tree-based models on tabular data~\cite{tree_vs_deepnn}, especially eXtreme Gradient Boosting (XGBoost)~\cite{chen2016xgboost} classifiers. These models do not support a 2-dimensional feature set so we collapse the features along the temporal axis and normalize using statistics from training data as part of input data preparation to the SVC. For our application, a radial basis function kernel provides the best results for SVC and we use the scikit-learn~\cite{pedregosa2011scikit} implementation and hyper-parameter search.

\noindent\smallskip\textbf{ConvLSTM.} We develop a baseline sequential model consisting of two depth-wise convolution layers along the temporal axis followed by two-layers of Long-Short-Term-Memory (LSTM)~\cite{hochreiter1997long} units for learning the temporal dependencies. These embeddings are then given to two fully-connected layers with the last layer providing categorical probabilities for the three classes to be used for the final prediction.  

\noindent\smallskip\textbf{BCResNet(Broadcasting along the Feature axis).}This design was originally proposed by Kim et al. ~\cite{kim2021broadcasted} for efficient key-word spotting applications and has a very small footprint, hence the choice of baseline for our application. It consists of repeating residual units that project the features (originally MFCCs of audio and in our case combination of spectral and temporal features) to 1D and then broadcast back to temporal space. 
\begin{figure}[b]

    \begin{center}
        \includegraphics[width=\linewidth]{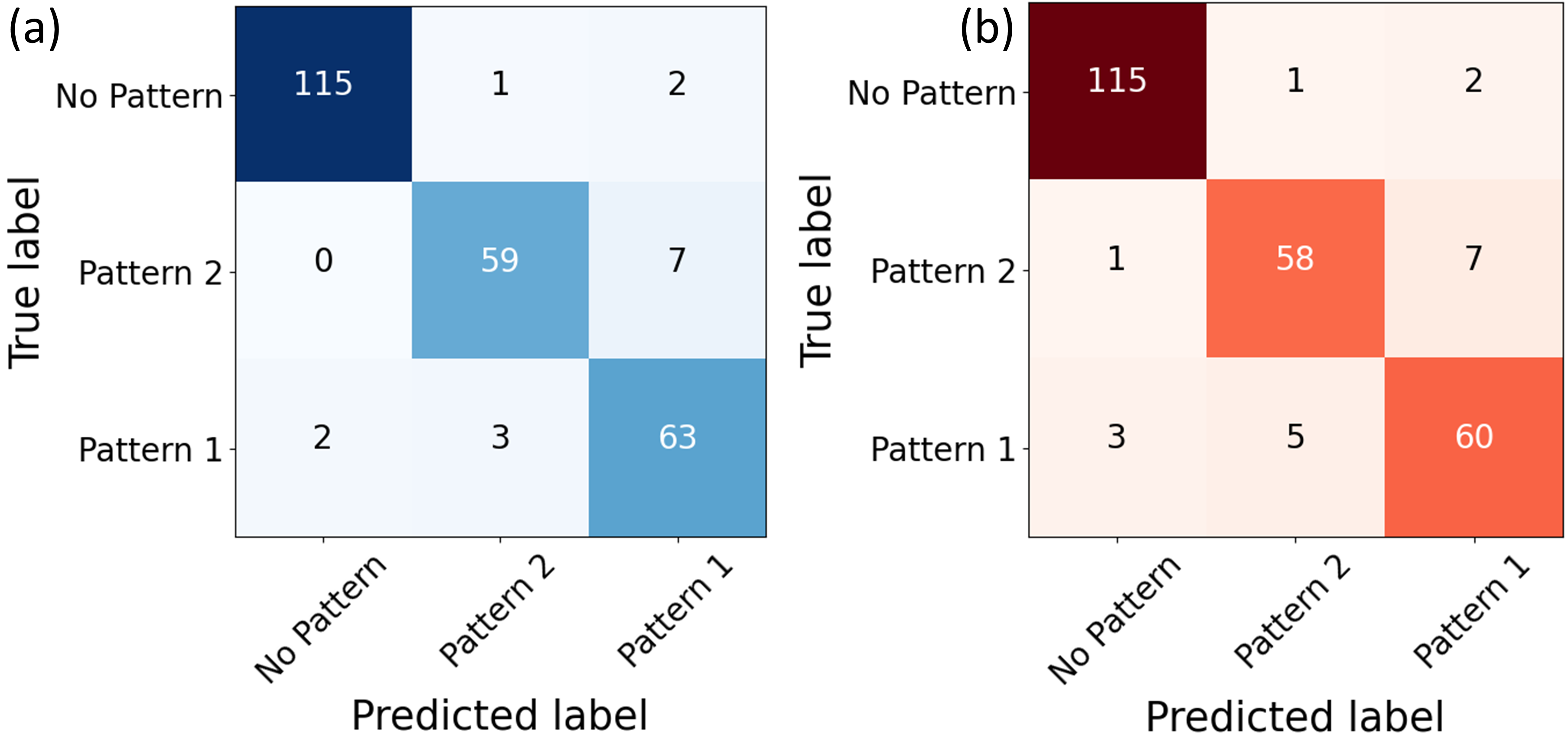}
        \caption{Confusion Matrix for \method{} network on (a) clean test samples (0.93 balanced accuracy) and (b) noisy test samples (0.91 balanced accuracy) from non-overlapping participants from the training data for No  Patterns, Pattern 1 (single teeth click) and Pattern 2 (double teeth click).}
        \vspace{-15pt}
        \label{fig:cm}
    \end{center}

\end{figure}

\noindent\smallskip\textbf{Broadcasting along the Temporal axis (\method{} Predictor Network).} This is a curated version broadcasting with residual connections for our application as described in detail in Section~\ref{subsubsec:predictornw} and Figure~\ref{fig:algoDesign}. The key updates are - introduction of layer normalization as the first transformation for input and using 1D temporal embeddings broadcasted to feature-wise 2D embeddings instead of vice-versa. We also evaluate a variant of \method{}'s predictor network where we pass the output of the temporal pooling layer through a convolutional self-attention layer to obtain an attention map. The attention map is used to scale the 1D vector to provide contextual importance to each feature. We use 2D convolution layers as the learnable parameters for attention and following the operations proposed by Vaswani et al.~\cite{vaswani2017attention}.



\begin{figure}
    \begin{center}
    \includegraphics[width=0.9\linewidth]{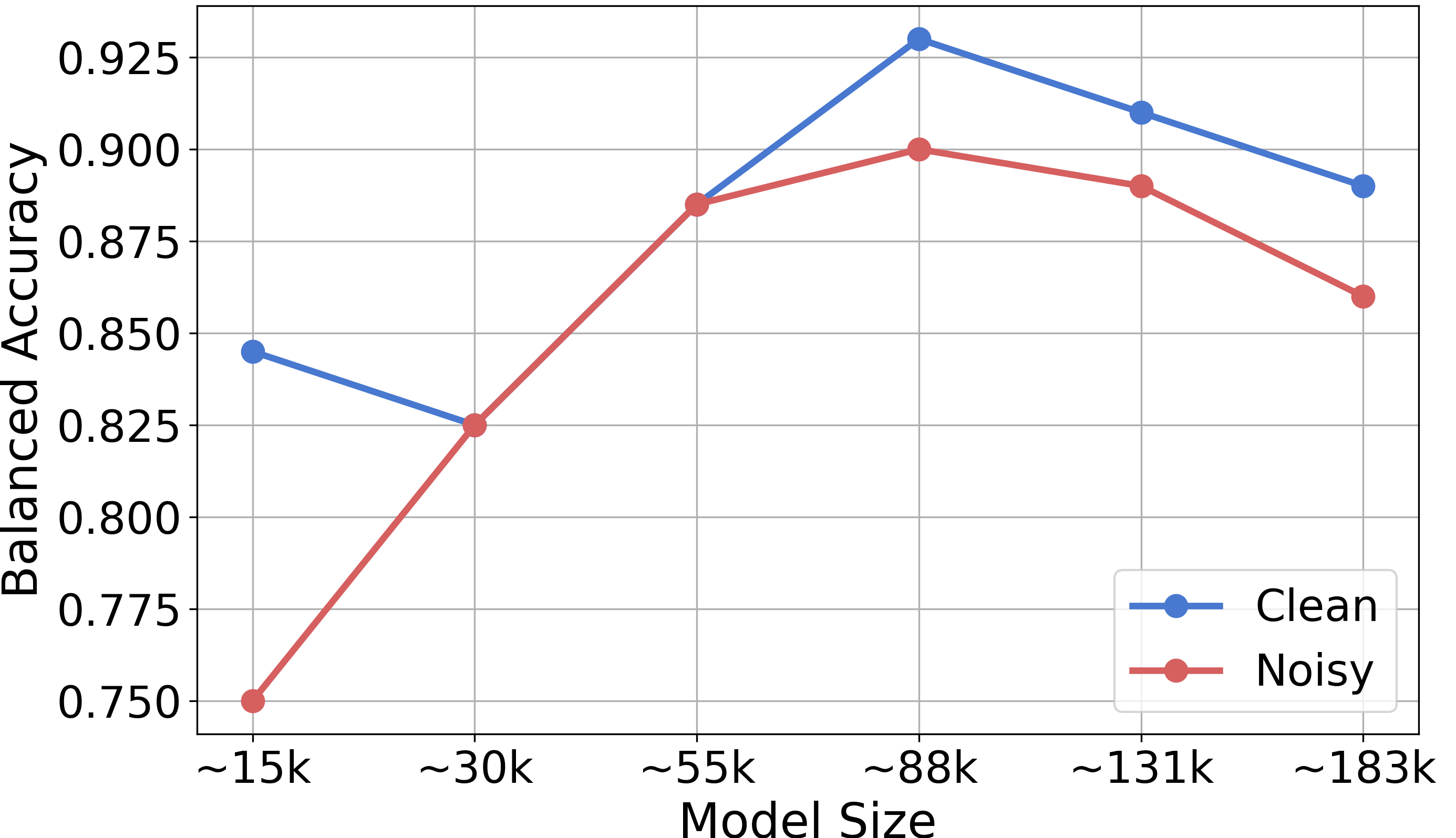}
    \caption{Impact of model size on \method{}'s predictor network performance. The model size is controlled by configurable channel dimensions.}
    \label{fig:model_size}
    \end{center}
\end{figure}

We summarize the results in Table~\ref{tab:summar_res} on clean and noisy test samples from non-overlapping subjects. \method{} outperforms all the models in the highest F1 scores per class and overall highest balanced accuracy of 0.93 and 0.91 on clean and noise test samples respectively. Figure~\ref{fig:cm} illustrates the confusion matrix for our proposed temporal broadcasting network in \method{} which has 7.14M Multiply and Accumulate (MMAC) units with approximately 88k trainable parameters. The temporal pooling and the instance normalization result in reducing the model size from the original BCResNet(Broadcasting along the Feature axis) by about 7k parameters. As discussed in Section~\ref{subsubsec:predictornw}, the configurable channel dimension across the recurring residual-broadcasting units helps control model size, we present a performance curve for various model sizes for \method{} in Figure~\ref{fig:model_size} for clean and noisy test data.

\subsection{Sensitivity Analyses} \label{subsec:feat_imp}
\begin{figure}
    \begin{center}
        \includegraphics[width=\linewidth]{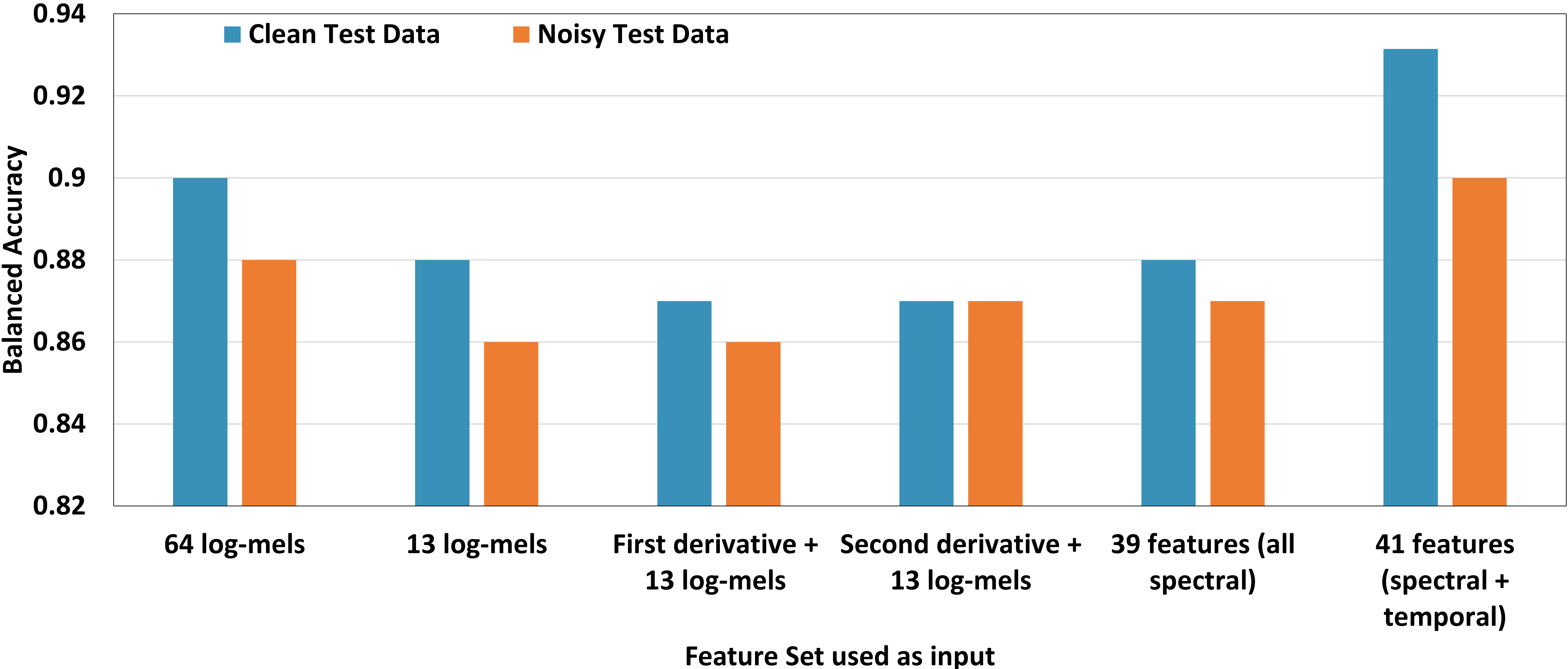}
        \caption{Using \method{}'s predictor network architecture to illustrate the importance of various features and the value in hand-crafting application-specific features in data-constrained settings.}
        \label{fig:feat_imp}
    \end{center}
\end{figure}

Generally, most large deep-learning networks are capable of modeling highly non-linear systems, and modest to no effort on feature engineering is required. However, in applications with very little training data such an assumption does not carry out well as the model sizes are conservative. We show that merely choosing log-mel features, even at a higher resolution, does not boost the model's performance as tailoring the feature set does. In our application, we are interested in capturing impulsive occlusal discreet dental events. Using derivatives of the spectral feature set (first and second derivatives of log-mel features). From Figure~\ref{fig:feat_imp} we can observe that although derived features from the original spectral features provide between 1-2\% improvement, it is really the combination of spectral(log-mel features and its first and second derivative) and temporal (short-term energy and zero crossing rate) features (41 features, last group in Figure~\ref{fig:feat_imp}) that boost the performance ~6\% over the use of only 13 log-mel features (second group in Figure~\ref{fig:feat_imp}). \textbf{Even if we increase the frequency resolution to 64 log-mels, the 41-dimensional spectral-temporal feature-set still outperforms it by ~3\%. This highlights the value in feature engineering for small-footprint networks like ours.} We present more details on the feature engineering in Section~\ref{subsubsec:feat_extraction}.

\subsection{Model Robustness}
\begin{figure}
    \begin{center}
        \includegraphics[width=\linewidth]{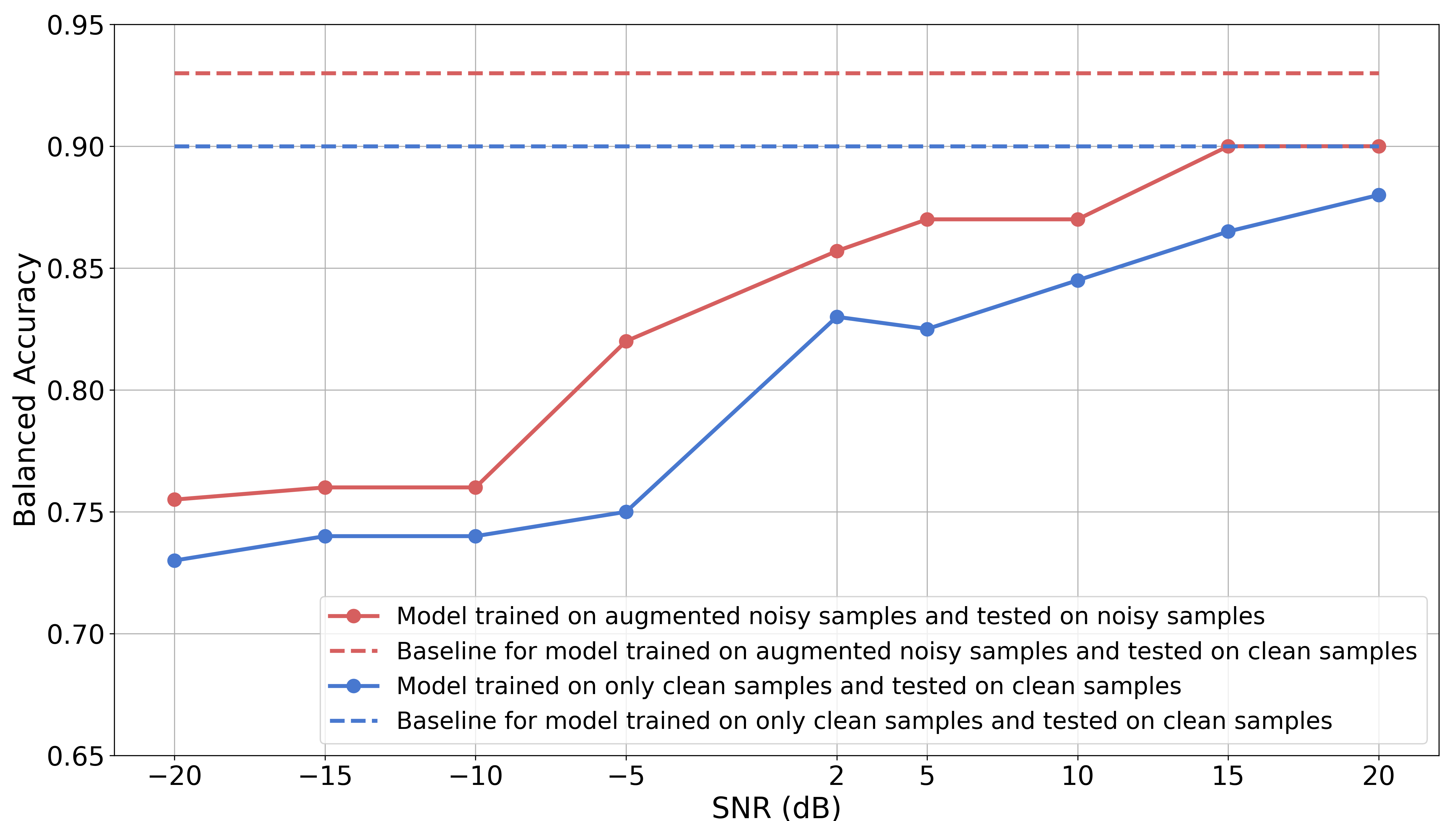}
        \caption{Evaluating the performance of two training schema - trained on clean samples only vs. trained using data augmentation strategy. The test samples are made noisy based on the characterization of the impact of noise floor on the accelerometers used for signal acquisition (refer Section~\ref{subsubsec:aug}). }
        \label{fig:robustness}
    \end{center}
\end{figure}

The emphasis of our work has been to develop a predictor network for detecting discreet oral signatures picked up by accelerometers that can be deployed in the real world. One of the key challenges to ensuring the model performs well in the wild is its robustness to noisy samples. As characterized in Section~\ref{subsubsec:aug}, background noise and motion artifacts may impact the signal of interest up to -23dB. To test the augmentation strategy employed we evaluate two models with exactly the same specification but one trained only on clean samples collected from subjects and the other using the data augmentation strategy detailed in Section~\ref{subsubsec:aug}. Then we use the best model checkpoints from both the training schemes and test on two types of data: 1) clean samples from unseen participants and 2) noisy samples from the same unseen subjects created by mixing acoustic and motion artifact-induced noise in a controlled way to log the model performance. Figure~\ref{fig:robustness} shows that the baseline for the model trained on clean sample points and tested on clean samples (from non-overlapping users) performs slightly worse (~2\%) than the model trained on augmented samples under the same settings. This could indicate overall better representations being learned when some noisy samples are introduced. \textbf{Altogether the model trained on augmented noisy samples performs on an average ~5\% better than the non-augmented training scheme under all the noise floor settings as illustrated in Figure~\ref{fig:robustness}.}

It is also noteworthy that we adopt an evaluation strategy for all metrics under the setting of leaving out 10\% of the participants and reporting the average score after testing on the samples from these non-overlapping users. This means for the chosen test participants samples that are user-specific like speech samples from the no pattern class, single click (pattern 1), and double click (pattern 2) events are left out from training and used for testing. This ensures the model's robustness to unseen participants (since we have seen evidence of user discriminability within samples in Section~\ref{subsub:feasibility}). 



\subsection{Visualization of other Discreet Oral gestures}\label{subsub:feasibility}
Given the limited precedent of previous works demonstrating the effectiveness in detecting and distinguishing occlusal patterns through smart glasses, our initial focus involves conducting a feasibility analysis for our proposed idea. To address several exploratory questions, we perform pilot analyses using data from three participants.

\noindent \textbf{Which discreet oral gestures can be picked up from accelerometers on smart glasses?}

\begin{figure}
    \begin{center}
        \includegraphics[width=\linewidth]{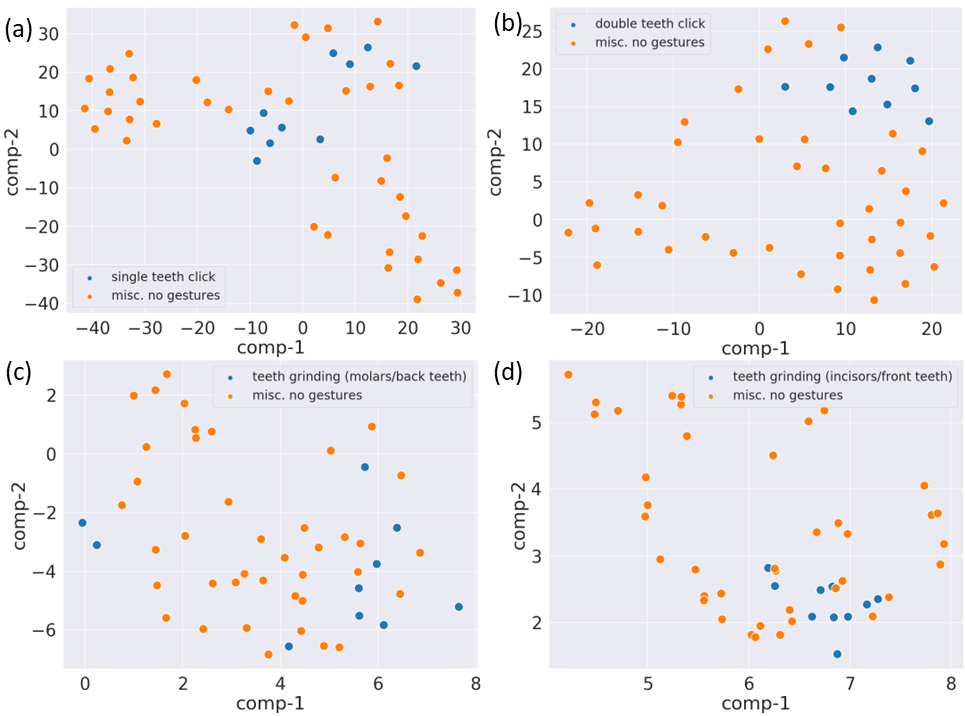}
        \caption{Visualization of various discreet oral gestures using t-SNE plots from 1 participant using 10 samples for each type. (a) single teeth click vs. no gesture scenarios,  (b) double teeth click vs. no gesture scenarios,  (c) teeth grinding using incisors (front teeth) vs. no gesture scenarios, (d) teeth grinding using molars (back teeth) vs. no gesture scenarios}
        \label{fig:eda_patterns}
    \end{center}
\end{figure}

\noindent The central feature to this technology is to command smart glasses discreetly in a hands-free manner. At the initial stage of the design we study various discreet oral gestures that can be picked up by an accelerometer placed at the nos epads of the glasses. We recorded 1-second gesture data where a participant issued four types of gestures 10 times. The gestures are 1) single teeth click, 2) double teeth click, 3) teeth grinding using the incisors (the front teeth) and 4) teeth grinding using the molars (back teeth). This technology is aimed at consumer products where highly prescriptive gestures are not warranted. Hence, in our studies users may use any combination of teeth from the upper and lower jaws or left or right sides of jaws to issue teeth clicks as convenient. In this preliminary study we transformed our 1 second temporal data to spectral features to give 13 log-mel~\cite{volkmann1937scale}. We then use a visualization tool, t-distributed Stochastic Neighbor Embedding (t-SNE)~\cite{van2008visualizing}, for reducing the dimensionality to a two-dimensional space as shown in Figure~\ref{fig:eda_patterns}. Our objective in this study is to assert a degree of distinguishability for the oral gestures. We also collect data from the same participants for various non-gesture events like drinking water, chewing, typing on the keyboard, silence, hearing music playing back from the smart glasses, etc. These are also 1s long and shown as orange scatter plots in Figure~\ref{fig:eda_patterns}. We observe better clustering ability of the \textit{teeth-click} (refer Figure~\ref{fig:eda_patterns}(a), (b)) like gestures. This could be attributed to their impulsive and distinct profile that is not usually encountered in other activities. However, teeth grinding can be very similar to mastication during eating food/drinks. Hence, we adopt teeth-clicking gestures as the choice of discreet non-verbal cues for smart glasses in our design.

\subsection{Field Test}\label{subsec:inthewild}
We design a real-time application where the participant can control the playback of music by clicking their teeth once or twice. We aim to assess the quality of this technology on two main axes - 1) adoption and 2) user experience. 



80.7\% of participants provided a mean rating of 3.74 (out of 5) for the adoption of this technology over the conventional means of controlling smart glasses. 96\% participants prefer having robustness to false positives and do not mind some false negatives. In addition, we subjectively gather a mean opinion score~\cite{pinson2012influence} for our technology. Since we cannot obtain a ground-truth reference we rely on direct feedback from users to indicate the performance of our method. We obtained a mean score of 3.33 and more than 88\% users reported performance of more than 3 on a scale of 1 to 5 (with 5 being the highest rating)\footnote{Specific numerical data from the user study analysis has been excluded due to proprietary restrictions. However, relative scores are provided to certify the end-to-end effectiveness of \method{}.}. Such positive user-study results indicate a reasonable rate of adoption of our proposed technology. Our ability to perform effectively in real-world scenarios, involving participants and categorical data that were previously unseen, instills confidence in the imminent mainstream adoption of this technology.

\section{Discussion, Limitations and Future Work}
In this work, we address the problem of hands-free non-verbal control for smart glasses. Such a technology is beneficial for users with limited limb functionality or speech disfluency beyond providing an immersive experience to augmented/virtual reality applications. While our demonstration overcomes the suboptimality of explicit gestures to control smart glasses, we acknowledge that there are areas for improvement that pique our interest for future exploration.

\noindent\textit{User Authentication.} Our preliminary experiments in Section~\ref{subsub:feasibility} provides evidence of using our technology and instrumentation to capture person-specific signatures with an accuracy 0.94 on a small scale dataset. This germinates its usage in user authentication~\cite{xie2022teethpass} applications. However, our real-world survey in Section~\ref{subsec:inthewild} indicates an apprehension by originally enthusiastic users of this technology for user authentication or other \textit{sensitive} applications like authorizing financial transactions by verifying user-identity from teeth click. Such a technology is more likely to be adopted if used for control functionality, in its current state. In future, we hope to explore more privacy-preserving techniques like federated learning~\cite{mcmahan2017communication} and iterate with user studies.

\noindent\textit{Personalization.} Currently, we use a universal prototype to collect data and conduct user studies, however, a personalized snug fit can improve the quality of data and hence the performance. On the algorithmic front, so far our focus has been to improve the upper-bound performance of a generalizable and robust model across individuals. In the future, we want to adopt strategies like few-shot learning~\cite{wang2022active} to enable personalization on the device for a user. Moreover, we would like to extend our setting as an open-set problem where the user can register new patterns with only a few sample examples by supporting class incremental learning schemes~\cite{wang2021few, koh2020incremental}. 

One of our current limitations is that self-speech robustness is tested only for the English language. There might be instances of some languages that naturally use more impulsive teeth-click-like gestures in conversation. Our current trained system may not work very well in such scenarios. 
Additionally, we aim to enhance our system's inclusivity by specifically addressing the needs of users with speech disabilities or health conditions such as bruxism, which may result in involuntary clicks. This will be achieved through conducting formal clinical user studies involving patients diagnosed with these conditions, using our system prototype.

\section{Conclusion}
An unobtrusive and discreet control expands the purview of smart glasses to seamlessly integrate without any social hindrance. We demonstrate a novel system instrumented to pick up discreet oral gestures using accelerometers on the nose pad of smart glasses. We develop a lightweight neural network robust to noise and variability across individuals due to dental or skull anatomy and successfully identify two types of teeth-clicking gestures with a balanced accuracy of 0.93. To transition our offline performance seamlessly into real-time in-the-wild applications, we purposefully craft a resilient model. Through a comprehensive exploration of application-specific augmentation techniques, we characterize the specifics of our problem. We further showcase the real-time version of this system to participants, receiving positive affirmation regarding the adoption and qualitative accuracy of the current prototype in the future.
\bibliographystyle{ACM-Reference-Format}
\bibliography{sample-base}










\end{document}